# High-resolution simulation of deep pencil beam surveys

Arno G. Weiß    Thomas Buchert

Max-Planck-Institut für Astrophysik, Garching, Germany.

**Abstract**

We report on a recent paper (Weiß & Buchert 1993), where we carry out pencil beam constructions in a high-resolution simulation of the large-scale structure of galaxies. As an example we present the results for the case of "Hot-Dark-Matter" (HDM) initial conditions (with scale-free $n = 1$ power index on large scales and $\Omega = 1$) as a representative of models with sufficient large-scale power. We use an analytic approximation for particle trajectories of a self-gravitating dust continuum and apply a local dynamical biasing of volume elements to identify luminous matter in the model. Using this method, we are able to resolve formally a simulation box of $1200h^{-1}$Mpc (e.g. for HDM initial conditions) down to the scale of galactic halos using $2160^3$ particles. Pencil beam probes are taken for a given epoch using the parameters of observed beams. In particular, our analysis concentrates on the detection of a quasi-periodicity in the beam probes. The simulation is designed for application to parameter studies which prepare future observational projects.

We find that a large percentage of the beams shows quasi-periodicities with periods which cluster at a certain length scale. The periods found range between one and eight times the cutoff length in the initial fluctuation spectrum. At significance levels similar to those of the data of Broadhurst *et al.* (1990), we find about 15% of the pencil beams to show periodicities, about 30% of which are around the mean separation of rich clusters, while the distribution of periodicity scales reaches values of more than $200h^{-1}$Mpc.

## 1 Introduction

Broadhurst *et al.* (1990) have reported an apparent quasi-periodicity of $128h^{-1}$Mpc in the combined data from pencil beams of the Kitt Peak and the Anglo-Australian Telescope (AAT) surveys (Broadhurst *et al.* 1988, Koo & Kron 1987). The combined pencil beam in total is $2200h^{-1}$Mpc deep reaching back to a redshift of $z \approx 0.5$, which probably probes a 'fair' depth of spacetime. The limitations in depth of present all-sky surveys indicate that we might learn more from the analysis of a sufficient number of deep pencil beams than from shallow all-sky surveys, at least as far as the understanding of the large-scale structure is concerned. (For a complete discussion of recent works see Dekel *et al.* 1992, Weiß & Buchert 1993.)

When simulating deep pencil beam observations in the context of a model for the formation of large-scale structure, the following requirements have to be met by the model: 1. It should mirror a 'fair' sample of the universe in the framework of some cosmogony, i.e., the simulation box should exceed the 'fair sample scale' which for positive spectral indices is the case, if the box covers the whole length of deep pencil beams, however, at least $250h^{-1}$Mpc for HDM initial conditions (see Buchert & Martínez 1993 for a definition of the 'fair sample scale'). As a minimal requirement we consider the box-length to be of the order of $1000h^{-1}$Mpc, so that periodic boundary conditions of the box produce no artificial signals. 2. At the same time, the model should resolve (at least formally) galactic scales in the whole box. This is important, since the beam probes are fairly narrow; results are only reliable, if the structure on the scale of typical cross-sections of the beams is resolved. In general, numerical simulations clearly do not meet these requirements.

We propose here a model which has several advantages for the purpose of modeling deep pencil beams. Our analytical model meets all the requirements mentioned above and overcomes the disadvantages of storage and CPU time intensive numerical simulations. In particular, the formal resolution of scales $\leq 0.5 h^{-1}$Mpc in the whole box, which is a sensible scale for a system galaxy + halo, is possible.

## 2  The model

### 2.1  The initial conditions

We generate a random Gaussian scalar field of fluctuations $S(\vec{X})$ on a $180^3$ Lagrangian mesh. According to HDM initial conditions (Bond & Szalay 1983) we take the power spectrum of the initial field to be of power-law form with a power index of $n = +1$ and a Gauss-smoothed cutoff $k_{\max}$. We identify this high-frequency cutoff $\lambda_{\min} = \frac{2\pi}{k_{\max}}$ with a large-scale characteristic length (for HDM: $\lambda_{min} = 13 \Omega_\nu^{-1} h^{-2}$ Mpc, where $\Omega_\nu$ is the density of one neutrino species relative to a flat background density, and $h$ the Hubble-constant in terms of 100 km/Mpcs). We assume a flat universe where baryonic matter can be neglected in comparison with the neutrino matter and therefore work with $\Omega_\nu = \Omega_{\text{tot}} = 1$. We express all length scales as multiples of the cutoff length $\lambda_{\min}$. Since we only use absolute lengths to specify the luminosity function in redshift space, we set $\lambda_{\min} = 25 h^{-1}$Mpc.

We realize the spectrum on a large box with periodic boundaries at $\lambda_{\max} = \frac{2\pi}{k_{\min}} = l_{\max} \cdot \lambda_{\min}$, where $l_{\max}$ should be large. Here we use $l_{\max} = 48$ which allows us to model large beams, like the two combined by Broadhurst *et al.*, according to our normalization, see *Fig.1*.

### 2.2  The evolution model

The "Zel'dovich-approximation" (Buchert 1989, 1992 and ref. therein) has proved to incorporate all aspects necessary for the present purpose down to the non-linearity scale of the perturbations and until a stage which could be roughly characterized by $\sigma = 1$ (Coles *et al.* 1993). A shortcoming of the "Zel'dovich-approximation" is the free shell-crossing resulting in "walls", which are too thick compared to numerical simulations. We consider this shortcoming to compensate the effect of the transformation to redshift space, which we neglect in the present work. (Tests on this compensation have been performed in a similar two-dimensional simulation by Buchert & Mo 1991). Also, the small scales are only formally resolved here, the action of gravity on these scales is not modeled properly. Therefore, if the *internal* structure of the pancakes matters, this scheme cannot be applied.

For this work, we use the first-order solutions of the general Lagrangian perturbation solutions (Buchert 1992) and restrict the general model, which involves two initial fluctuation fields (given at $z_0 = 1000$), one for the peculiar-velocity $\vec{U}(\vec{X})$ and one for the peculiar-acceleration $\vec{W}(\vec{X})$, by assuming irrotationality of the initial velocity field. Also, we assume $\vec{U}(\vec{X}) = \vec{W}(\vec{X}) t_0$, i.e., both fields are spatially congruent. With this restriction, the general first order mapping from Lagrangian coordinates $\vec{X}$ to comoving Eulerian coordinates $\vec{q}$ practically reduces to Zel'dovich's mapping and reads (for a flat background universe):

$$\vec{q} = \vec{F}(\vec{X}, z) = \vec{X} + \left( \left( \frac{1+z_0}{1+z} \right) - 1 \right) \frac{3}{2} \nabla_X S^{(1)}(\vec{X}), \qquad (1)$$

$$\Delta_X S^{(1)}(\vec{X}) = \text{trace}\left( \frac{\partial^2 S}{\partial X_i \partial X_j} \right) t_0,$$

where $S$ is the peculiar velocity potential:

$$\vec{U} =: \nabla_X S.$$

This mapping depends non-locally on the initial condition $S$, in contrast to Zel'dovich's mapping. However, in the first-order case we can make it local by setting $\nabla_X S^{(1)}$ to $\nabla_X S$ without loss of generality (Buchert 1992).

In a next step we shall incorporate the second-order solution (Buchert & Ehlers 1993, Buchert 1993), which extends the range of validity of the model to later times and also to smaller scales.

## 2.3 Gigaparsec realizations at high resolution

In order to establish the link between 'fair' and galactic scales, we have to perform simulations of very high spatial resolution. The — at least from the viewpoint of simulating large samples of the universe — main advantage of the analytical model (1) is the possibility of improving particle number and hence resolution of the simulation by interpolation of the mapping (1). For this, we compute the generating field $\vec{U}(\vec{X})$ on a grid of $180^3$ points, where the resolution of the grid must allow for a sufficiently accurate resolution of the highest frequency component in the spectrum. Then we interpolate trilinearly between the grid points, giving us an easy way of adapting the number of particles to the needs of our simulation. This way, we can resolve our simulation volume of $1200^3 h^{-3} \mathrm{Mpc}^3$ into $2160^3$ particles, giving a linear resolution of $555 h^{-1} \mathrm{kpc}$, which is a sensible scale for a system of galaxy + halo for the simulated mass limit. (The interpolation method as a tool for large-scale structure modeling has been introduced and illustrated by Buchert & Bartelmann (1991)). The resolution used ($2160^3$) is extremely high compared with numerical standards, the realization of a box is comparatively fast. To our experience, the chosen resolution represents the lowest resolution necessary for a sensible simulation of deep pencil beam data. This is because the model has to cover at least 2 times three orders of magnitude in spatial scale. Thus, the model is sensible also to small-scale clustering properties of galaxies within walls, which is important for the probability of a beam to hit the wall (compare Ramella *et al.* 1992). In this line the model can be improved. One important step of improvement, which we already mentioned, is the implementation of a second-order correction into the displacement mapping. The second-order approximation models the tidal effect of self-gravity on the relevant spatial and temporal range. Of course, the simulation of the merging process of galaxies (here modeled simply by random selection, see the next subsection) as well as hydrodynamics will be necessary to verify our phenomenological selection criteria to identify luminous matter.

## 2.4 Local dynamical biasing algorithm

As explained in more detail in (Buchert 1991a,b), we use a local (non-linear) density bias to determine a subfraction of luminous matter in the model. This biasing method is *dynamical* in the following sense: We follow volume elements of *comoving* size $\alpha$ along their trajectories given by the analytical mapping (1). We then consider the density at the element. We call the element *luminous* all the time after it has reached a threshold $\chi_c$ in the Lagrangian density excess:

$$\chi(\vec{X}, z) := \frac{\varrho}{\varrho_b} = \frac{1 + \delta_0(\vec{X})}{\det\left(\frac{\partial F_i}{\partial X_j}\right)(\vec{X}, z)}. \tag{2}$$

On the resolved scale, a lower bound on the masses of objects $M_c$ in the simulation can be assumed, the corresponding density threshold value can be calculated according to:

$$\chi_c = \frac{M_c(z=0)}{\varrho_b(z=0)\alpha}, \tag{3}$$

where the resolved (comoving) volume is $\alpha = (555 h^{-1} \mathrm{kpc})^3$, related to physical volumes $\Delta V$ as $\alpha = \Delta V(1+z)^3$. If the resolved volumes are equal for all $\vec{X}$ and if $\chi = const.$ for all $z$, we find that everywhere and at any epoch the physical masses produced are equal. Thus, for e.g. $\chi_c = 3.0$, all points in the simulation represent luminous matter above a lower mass bound $M_c = 1.42 \cdot 10^{11} h^{-2} M_\odot$, or luminosity limit, respectively. Note that the choice of *comoving* volumes on which the threshold operates compensates for the enhanced background density at earlier epochs.

We then select randomly (*after* thresholding) a subsample of potentially luminous volume elements such that overall $\Omega_{\rm lum} = 0.01$. This random selection can be viewed to mimic merging of luminous volume elements (elementary masses) into larger objects.

Clearly, this selection criterion does not relate the large-scale density field directly to the formation conditions of individual galaxies. We consider this selection as a phenomenological link to the environmental conditions of galaxy formation considerably improving the standard Eulerian biasing scheme (compare Kates *et al.* 1991).

Since we have discriminated luminous and dark matter, we have introduced a non-standard HDM model. As the present local biasing algorithm is non-linear we can have interesting consequences for the local structure characteristics (compare Buchert & Martínez 1993).

## 3 Simulation of pencil beam surveys

We restrict our simulations to a cubic box of $1200 h^{-1}$Mpc base length. Starting in one corner of the cube (the observer's position) we randomly draw pencil beams of solid angles of $20'$ reaching into the cubes volume. This way, the size of the simulated beams is still large enough to expect about 10 periods, which is enough for a reasonable statistics of periods. We conctruct three ensembles of pencil beams, each consisting of 100 beams. The difference between the ensembles is due to a different realization of the model, i.e., a different random set of amplitudes and phases.

### 3.1 Selection of a beam catalog from the simulated point process

Pencil beam galaxy catalogues are extracted from the model in four steps:

1. Geometry: All particles of the simulation which lie in the survey geometry (in case of this simulation, a narrow cone of $20'$ solid angle extending out to $1200 h^{-1}$Mpc from one corner of the cubic simulation volume) are extracted.

2. The threshold scheme: All particles whose Lagrangian density excess exceeds a critical value $\chi_c$ are selected as belonging to "potentially" luminous matter. We choose $\chi_c = 3.0$.

3. Overall matter density: We reduce the number of particles within our survey geometry such that a value of $\Omega_{\rm lum} = 0.01$ results by randomly selecting particles (*after* thresholding). Here we have to account for the limited mass scale of our simulation (only masses $\geq M_c(\chi_c)$ are simulated) by setting the selection probability of the particles to a correspondingly lower value. After this step the particles can be considered to represent galaxies. The resulting mean density of galaxies (in the simulated mass range) is $4.2 \times 10^{-3} h^3 {\rm Mpc}^{-3}$. (A slice through our simulation volume corresponding to this stage can be seen in Fig. 1)

4. Visibility: We assign to each galaxy a detection probability (i.e. a probability that its luminosity lies in the range of visible values, depending on its distance from the observer and on the simulated survey limit $m_{\rm lim}$). We use a Schechter-type luminosity function with parameters according to Broadhurst *et al.* (1990). Taking the limited range of masses (and therefore luminosities) simulated in our model, the resulting mean luminosity density of $\mathcal{L} = 8.4 \times 10^7 L_\odot {\rm Mpc}^{-3}$ (for $h = 0.5$) is in good agreement with the estimate of $\mathcal{L} \sim 10^8 L_\odot {\rm Mpc}^{-3}$ given by Kirshner *et al.* (1979).

After the last step in this procedure, the resulting set of particles can be considered as a catalogue of visible galaxies (for the simulated observer and survey limit). Of course, this scheme can easily be extended to other survey geometries.

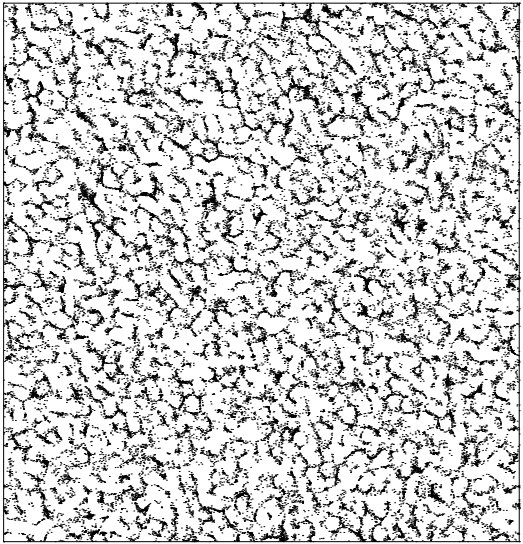

Figure 1: A slice of the simulation. Thickness is $1h^{-1}$Mpc with a side length of $1200h^{-1}$Mpc. Only galaxies according to $\Omega_{\rm lum} = 0.01$, as obtained by our selection scheme for a threshold of $\chi_c = 3.0$, have been included in the plot.

60817 galaxies

## 4  Results

The number of periodic pencil beams naturally depends on the criteria used to decide whether a beam is called periodic or not. Apart from visual impressions (of the distance distribution and pair counts) we concentrate on three criteria: The significance $X$ of peaks in the power spectrum (Szalay et al. 1991), given as peak-power to white-noise-power ratio (our main criterion), a trial period folding algorithm, and a method using the change of slope between left-sided and right-sided linear regressions of the ratio of unnormalized pair counts (Mo et al. 1992).

Requiring only a significance of $X \geq 6$, which is about half that of the beam of Broadhurst et al., we find as much as 41% of the beams periodic, while for $X \geq 11.8$ (the value found by Szalay et al.) we still have 15% of the beams periodic. The difference between different ensembles is in this case not extreme, as can be seen in Table 1.

| $X_p \geq$ | 14.0 | 12.0 | 11.8 | 10.0 | 8.0 | 6.0 |
|---|---|---|---|---|---|---|
| %(1) | 10.0 | 12.0 | 13.0 | 19.0 | 25.0 | 40.0 |
| %(2) | 7.0 | 13.0 | 13.0 | 19.0 | 28.0 | 44.0 |
| %(3) | 10.0 | 17.0 | 18.0 | 25.0 | 30.0 | 40.0 |
| %$\Sigma$ | 9.0 | 14.0 | 14.7 | 21.0 | 27.7 | 41.3 |

Table 1: The percentage of beams with at least one peak of $X \geq X_p$, for different minimal significance levels $X_p$.

The distribution of periods for significance levels above $X_p = 6.0$ and $X_p = 11.8$ is shown in Figure 2. For higher significances, the detected periods start slightly above the cutoff scale of the initial spectrum, while few periods of low significance can also be found slightly below this scale. About 50% of the periodic beams have periods in the interval $[\lambda_{\min}, 2\lambda_{\min}]$. Around typical values of rich cluster distances ($\lambda \in [50, 70\ ]h^{-1}$ Mpc for $\lambda_{\min} = 25h^{-1}$Mpc) we found $\simeq 30\%$ of the *periodic* pencil beams with significance $X \geq 11.8$. However, only a few beams with periods $\geq 100h^{-1}$Mpc are found. This could be due to the lack of small scale clustering inside the walls, which would favour the detection of large periods because many walls might be missed. In our realizations we didn't find evidence for the occurence of large periods due to this effect. Large periods are (in our simulation) mainly the results of the detection of clusters instead of walls in the pencil beam. The detection of large periods in this case is evident: firstly, due to a lower probability to hit a cluster instead of a wall, secondly, due to a higher significance of peaks in the power spectrum (the number of galaxies is larger in clusters).

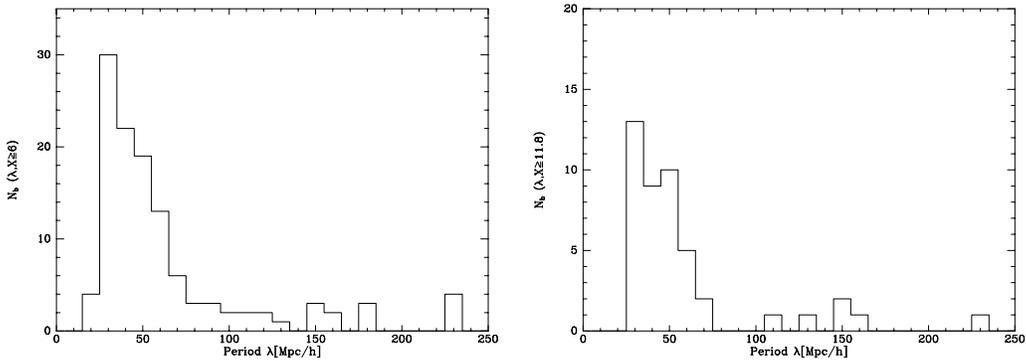

Figure 2: The distribution of periods for beams with a most significant peak in the power spectrum of at least $X = 6.0$ *(left)* and $X = 11.8$ *(right)*.

Figure 3 shows the distribution of galaxies, pair count histograms, change of slope diagrams and power spectra for four selected pencil beams, while Figure 4 shows the surface density of galaxies for two selected beams, computed *without* correction for the luminosity function. Usually, the peaks lie in the region of $1 - 2.5$ galaxies/$h^{-2}\text{Mpc}^2$, but single peaks can extend to 10 times higher densities. Obviously, the higher peaks are caused by the intersection of rich galaxy clusters with the beam, while the generic case of lower density peaks represents the intersection with walls.

## 5 Discussion and relation to other work

Compared to other work the relative abundance of 15% peaks of a significance of at least the level of the pencil beam observed by Broadhurst *et al.* (1990) detected in our simulations is quite high. The occurence of periodicities in pencil beams as extracted from a model with a truncated spectrum is easy to understand. Only van de Weygaert (1991) and Subbarao & Szalay (1992) give information about the occurence of periodicities in their Voronoi tesselation simulations. While van de Weygaert finds for generic spectra also about 15% periodic beams (without using a quantitative measure to define these periodicities), Subbarao & Szalay find similar rates of periodicities only for Voronoi tesselations using seeds on a cubic grid, and much less for Poissonian seeds. So, our model seems in fact to favour the detection of periodicities in pencil beams.

However, there remain some questions about the detected scale of the periods, since the rate of detection of large periods in our model is quite low, at least if we use standard HDM normalization. Statistical information to compare with is scarce. As long as the statistical status of the observations is not improved, a low abundance of high periods does not argue against this model. The detection of large periods may well be the result of small-scale clustering inside the walls, giving a relatively high probability for the pencil beam to miss walls (Ramella *et al.* 1992). Since we have used a simple (first-order) approximation, a random selection of biased volume elements generates a more or less Poissonian distribution of test particles over the sheets. Therefore, modeling the small-scale clustering inside of walls better (by using second-order approximation theory) might also change the abundance of large periods.

The surface densities of galaxies inside walls are around $1 - 2.5 h^2 \text{Mpc}^{-2}$ without a luminosity selection (see *Fig. 4*). These values agree well with those of $0.2 - 0.3 h^2 \text{Mpc}^{-2}$ found by de Lapparent *et al.* (1991) and Ramella *et al.* (1992) for the "Great Wall" in the CfA-survey, allowing for a factor of $\sim 4$ alone by the effect of the luminosity function on the apparent surface densities.

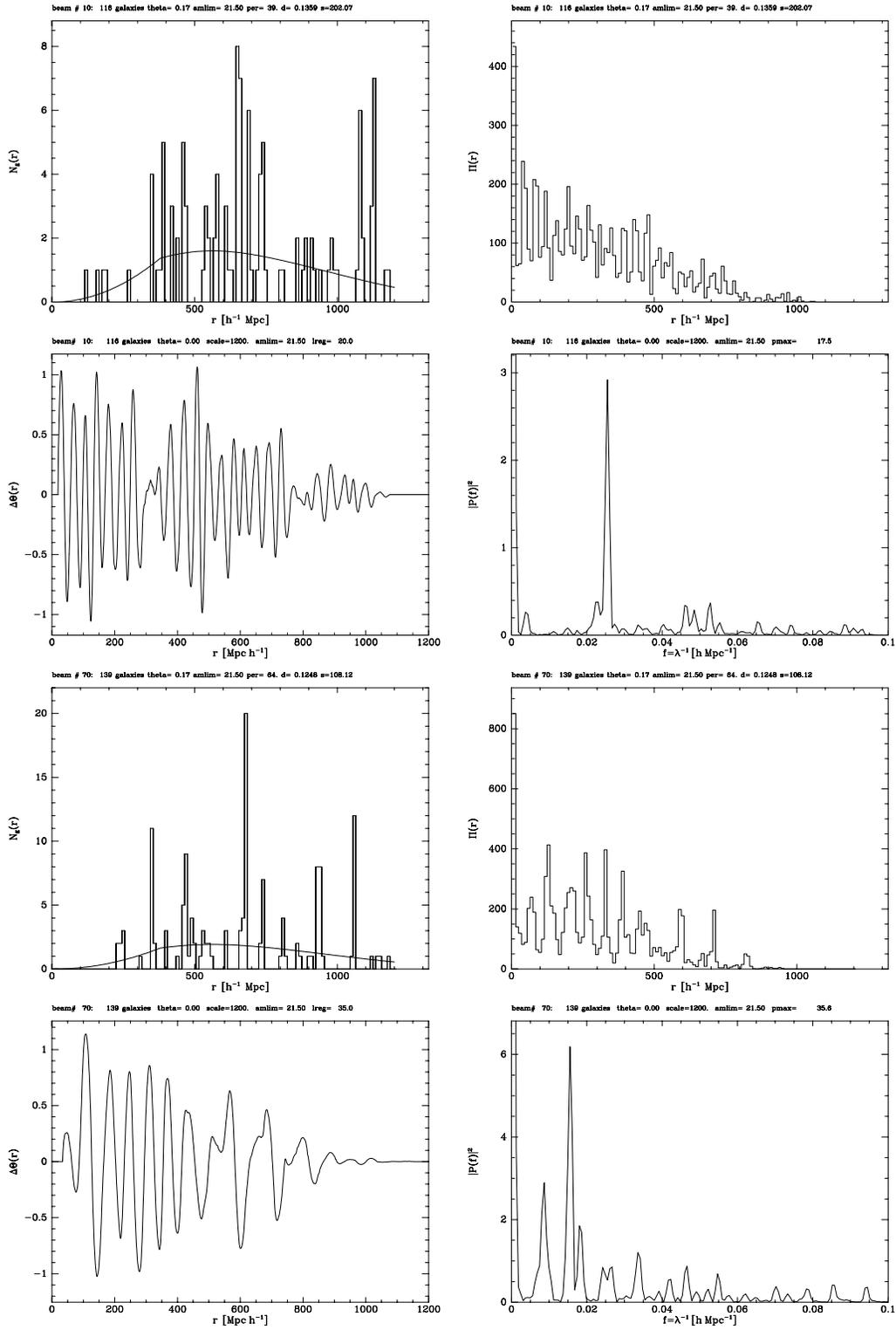

Figure 3: The distribution of galaxies, pair count histograms, change of slope diagrams and power spectra for two selected pencil beams. (The bin size is $10h^{-1}$Mpc, the significances of strongest periods are $> 11$.)

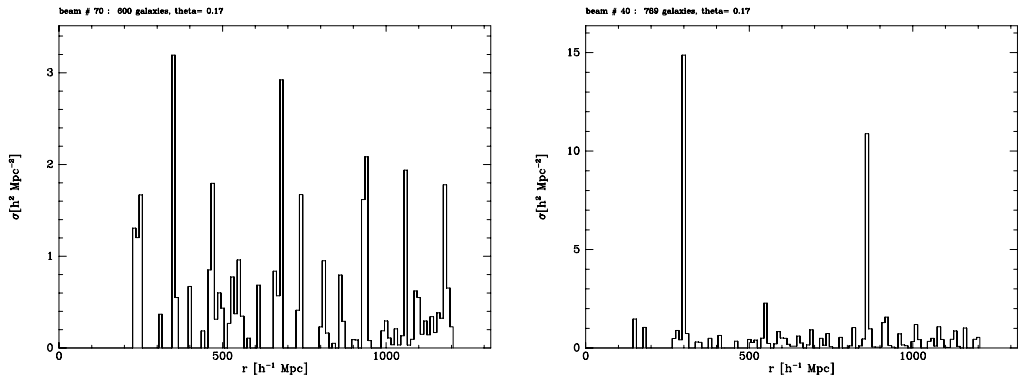

Figure 4: The surface density of galaxies for two selected beams, computed *without* correction for the luminosity function.

Compared to typical scales in the large-scale distribution of galaxies as reported by Mo *et al.* (1992), the maximum of the distribution of periodicities could naively be expected around the scale $60h^{-1}$Mpc, which is a common scale in a huge list of catalogues including Abell clusters, all-sky surveys and pencilbeam surveys. According to our normalization the maximum is identified with a scale of about $35h^{-1}$Mpc indicating that either the cutoff should be normalized to a larger value, or the power on large scales should be increased. However, we have done an independent check of the presence of this scale in the simulation by using the unnormalized paircount method for several $1h^{-1}$Mpc slices of the box. We found that indeed in some slices a clear periodicity of $\simeq 60h^{-1}$Mpc can be detected corresponding to the presence of rich clusters in these slices, while for most of the slices only the cutoff scale was present. If the dominance of a $60h^{-1}$Mpc scale is due to the presence of *rich* clusters, then our normalization is still adequate. However, in order to draw definite conclusions, a systematic study of typical scales in the galaxy and cluster distribution is indispensible, which lies beyond the scope of the present work.

In this line it is interesting to note that Mo *et al.* (1992) have also analyzed the pencil beam data of Broadhurst *et al.* They found a typical scale of $130h^{-1}$Mpc in agreement with other methods, which is due to a significant feature in the data. However, if a small field including this feature is excluded from the data, the period drops to $60h^{-1}$Mpc in agreement with the commonly traced scale of different objects. This suggests that large periods are indeed special, and the basic period should be expected around the $60h^{-1}$Mpc scale, which approximately corresponds to the average period (weighted with the number of periodic beams) for our normalization.

# Acknowledgements

*We would like to thank Robert Klaffl for letting us use his highly optimized 3D Zel'dovich approximation code and for technical help, Houjun Mo for providing us his pair count programs as well as for fruitful discussions, and Tom Broadhurst for explaining to us his trial period folding algorithm as well as for discussions.*
*Extensive remarks on the manuscript were made by Gerhard Börner.*
*AGW likes to thank Gerhard Börner for supervising his Diploma thesis on this topic.*
*TB acknowledges financial support by DFG (Deutsche Forschungsgemeinschaft).*